\documentclass[lettersize,journal]{IEEEtran}
\usepackage{cite}
\usepackage{amsmath,amssymb,amsfonts}
\usepackage{algorithmic}
\usepackage{graphicx}
\usepackage{textcomp}
\usepackage[table]{xcolor}
\usepackage{hyperref}
\usepackage{enumitem}
\usepackage{subfigure}
\usepackage{booktabs}
\usepackage{multirow}
\usepackage{makecell}
\usepackage{amsmath}
\usepackage[most]{tcolorbox}

\newtcolorbox{questionbox}{
  colback=gray!6,
  colframe=gray!20,
  arc=2mm,
  boxrule=0.5pt,
  left=1mm, right=1mm,
  top=1mm, bottom=1mm
}
\usepackage[normalem]{ulem}
\definecolor{LightBlue}{RGB}{233,243,250}
\definecolor{HeaderBlue}{RGB}{223,234,242}
\def\BibTeX{{\rm B\kern-.05em{\sc i\kern-.025em b}\kern-.08em
    T\kern-.1667em\lower.7ex\hbox{E}\kern-.125emX}}
\begin{document}

\DeclareRobustCommand{\loongrightarrow}{%
  \DOTSB\relbar\joinrel\relbar\joinrel\rightarrow
}
\DeclareRobustCommand{\loongmapsto}{\DOTSB\mapstochar\loongrightarrow}

\DeclareRobustCommand{\looongrightarrow}{%
  \DOTSB\relbar\joinrel\relbar\joinrel\relbar\joinrel\rightarrow
}
\DeclareRobustCommand{\looongmapsto}{\DOTSB\mapstochar\looongrightarrow}

\title{3D Gaussian Splatting: A New Paradigm for Wireless Environment Representation}
\author{Jinya Zhang, Jiajia Guo, \IEEEmembership{\normalsize{Member,~IEEE}},~Chao-Kai~Wen, \IEEEmembership{\normalsize{Fellow,~IEEE}}, and Shi~Jin, \IEEEmembership{\normalsize{Fellow,~IEEE}}

}

\maketitle

\begin{abstract}
Wireless channels are shaped by the surrounding propagation environment, making explicit environment representation an effective approach for characterizing channels. By combining explicit environment modeling with learnable radio frequency characteristics, 3D Gaussian splatting (3DGS) emerges as a new paradigm for wireless environment representation. This article establishes a systematic perspective on 3DGS-enabled wireless communications in terms of environment representation. We first clarify the rationale for extending 3DGS from visual scene modeling to wireless communications and formulate a unified framework for wireless environment representation. Building on this framework, we further investigate the applicability of 3DGS-based representations to a broad range of communication and sensing tasks, and characterize key considerations for practical deployment. Finally, emerging directions are highlighted, including cross-scene generalization, near-field channel modeling, real-world data collection, and lifecycle management.
\end{abstract}

\section{Introduction}

The performance of wireless communication systems strongly depends on the availability of accurate channel state information (CSI)~\cite{cuitowards}. Toward future wireless networks, increasingly large antenna arrays and wider bandwidths are expected to support higher spectral efficiency and larger network capacity. However, these advances substantially increase the dimensionality of CSI and the burden of channel acquisition. Conventional CSI acquisition, which typically relies on pilot transmission, channel estimation, and CSI feedback, incurs rapidly growing overhead as the number of antennas and frequency resources increases. Therefore, acquiring accurate CSI with reduced overhead has become an increasingly important requirement for future wireless networks.

Since wireless channels are shaped by interactions with their surroundings, wireless environment representations aim to characterize channels in relation to the environment. Such representations, including radio maps and channel knowledge maps, can provide useful priors for CSI acquisition, thereby reducing overhead. To construct these representations, existing methods can be broadly classified into data-driven and physics-based approaches. Data-driven methods, particularly neural networks (NNs), can learn complex radio frequency (RF) patterns from channel measurements, but often rely heavily on data due to limited physical guidance. In contrast, physics-based methods, such as ray-tracing, explicitly model the propagation environment, offering interpretability while requiring meticulous modeling and high computational costs. These limitations raise a natural question:
\begin{questionbox}
\centering
\noindent {\textit{Can we establish a new paradigm for wireless environment representation that learns complex RF patterns constrained by explicit physical  structures?}}
\end{questionbox}

Recent advances in 3D Gaussian splatting (3DGS)~\cite{kerbl20233d} have introduced a new paradigm for wireless environment representation,  bridging explicit environment modeling with data-driven learning of complex RF characteristics. Originally developed in computer vision for novel view synthesis, 3DGS explicitly represents a scene with an ensemble of anisotropic 3D Gaussians. Each Gaussian can be intuitively viewed as a fuzzy ellipsoid with learnable geometric and appearance attributes. The former determine its position and shape, while the latter characterize the observed light. By aggregating the contributions of all Gaussians, 3DGS efficiently synthesizes images from novel viewpoints. Notably, the 3rd Generation Partnership Project has examined 3DGS as a visual media representation for compression and transmission, reflecting growing industry interest in this technology. The shared electromagnetic (EM) nature of light and radio waves motivates extending 3DGS from the optical to the RF domain. In the RF domain, Gaussians can be reinterpreted as virtual radiators with RF-specific attributes, forming a queryable wireless representation for predicting channel knowledge at unmeasured locations. Pioneering studies, including RF-3DGS~\cite{zhang2026rf} and WRF-GS~\cite{wen2025wrf}, have demonstrated the potential of 3DGS for wireless environment representation.

This article establishes a systematic perspective on 3DGS as a new paradigm for wireless environment representation. It clarifies the rationale for extending 3DGS from vision to wireless communications, formulates a general framework for 3DGS-enabled wireless environment representation, examines 3DGS's role in downstream wireless applications, and distills the key considerations for practical deployment. It also identifies several promising yet underexplored directions for future research. The remainder of this article is structured as follows. The next section introduces 3DGS fundamentals. Section III establishes the general paradigm of 3DGS-enabled wireless environment representation. Section IV includes downstream wireless applications, while Section V addresses key issues for practical deployment. Sections VI and VII present future research directions and conclude the article, respectively.

\begin{figure*}
    \centering
    \includegraphics[width=1\linewidth]{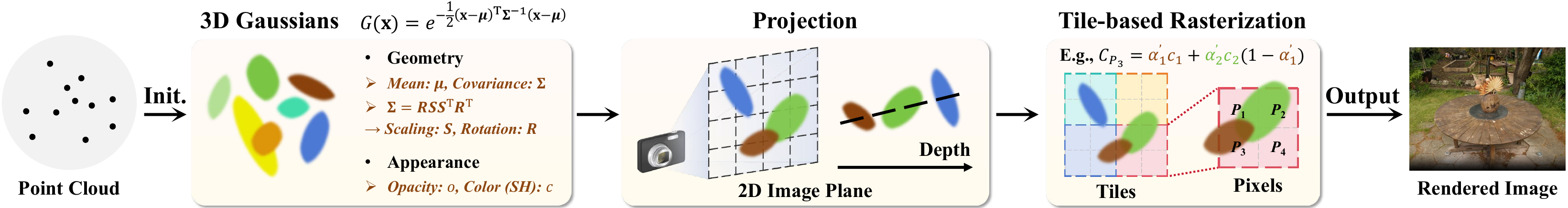}
    \caption{Overview of the 3D Gaussian Splatting pipeline. A point cloud initializes 3D Gaussian primitives with geometric and appearance attributes, which are projected onto the image plane and rendered via tile-based rasterization to synthesize an image from a specified viewpoint.}
    \label{fig:3DGS}
\end{figure*}

\section{3DGS Fundamentals:\\From Vision to Wireless Communications}
\subsection{3DGS Basics}
3DGS~\cite{kerbl20233d} is a computer vision technique for novel view synthesis that generally involves 3D scene reconstruction and rendering. \textit{3D scene reconstruction} aims to recover a digital 3D model from a collection of views. Based on the reconstructed scene, \textit{rendering} generates 2D images from queried camera viewpoints. As illustrated in Fig.~\ref{fig:3DGS}, the basic pipeline of 3DGS consists of three main components.

\subsubsection{3D Gaussian representation} Starting from an initial point cloud, 3DGS represents the scene using a set of anisotropic 3D Gaussian primitives. Each Gaussian is parameterized by a mean $\boldsymbol{\mu}$ and a covariance matrix $\boldsymbol{\Sigma}$, which determine its position and shape, respectively. To keep $\boldsymbol{\Sigma}$ positive semi-definite during optimization, the covariance matrix is decomposed into a scaling matrix $\mathbf{S}$ and a rotation matrix $\mathbf{R}$ as $\boldsymbol{\Sigma}=\mathbf{R}\mathbf{S}\mathbf{S}^{T}\mathbf{R}^{T}$. In addition to geometric attributes, each Gaussian is associated with an opacity $o$ and a view-dependent color $c$ represented by spherical harmonics (SH). 
\subsubsection{Projection} Given a specific camera viewpoint, the 3D Gaussians are projected onto the image plane as 2D Gaussians. First, their means and covariances are transformed from world coordinates to camera coordinates using the camera extrinsic parameters. The camera intrinsic parameters and perspective projection are then applied to map the transformed 3D Gaussians onto the 2D image plane. Consequently, the projected position and shape of each Gaussian depend on both its 3D geometry and the camera.
\subsubsection{Tile-based rasterization} Finally, the projected 2D Gaussians are efficiently rendered using a tile-based rasterizer. The image plane is divided into non-overlapping tiles to enable parallel processing on GPUs. For each tile, the overlapping Gaussians are identified and sorted by depth. Their contributions to each pixel are then accumulated through $\alpha$-blending, where closer Gaussians attenuate the contributions of farther ones, thereby accounting for occlusion. The resulting pixel values form the rendered image.

During training, the Gaussian parameters are optimized by minimizing the loss between the rendered and ground-truth images across multiple views. Meanwhile, adaptive density control periodically densifies or prunes the Gaussians to refine the scene representation. Once reconstructed, the scene enables efficient rendering from unseen viewpoints.

\subsection{From Vision to Wireless Communications}
\begin{figure*}
    \centering
    \includegraphics[width=0.9\linewidth]{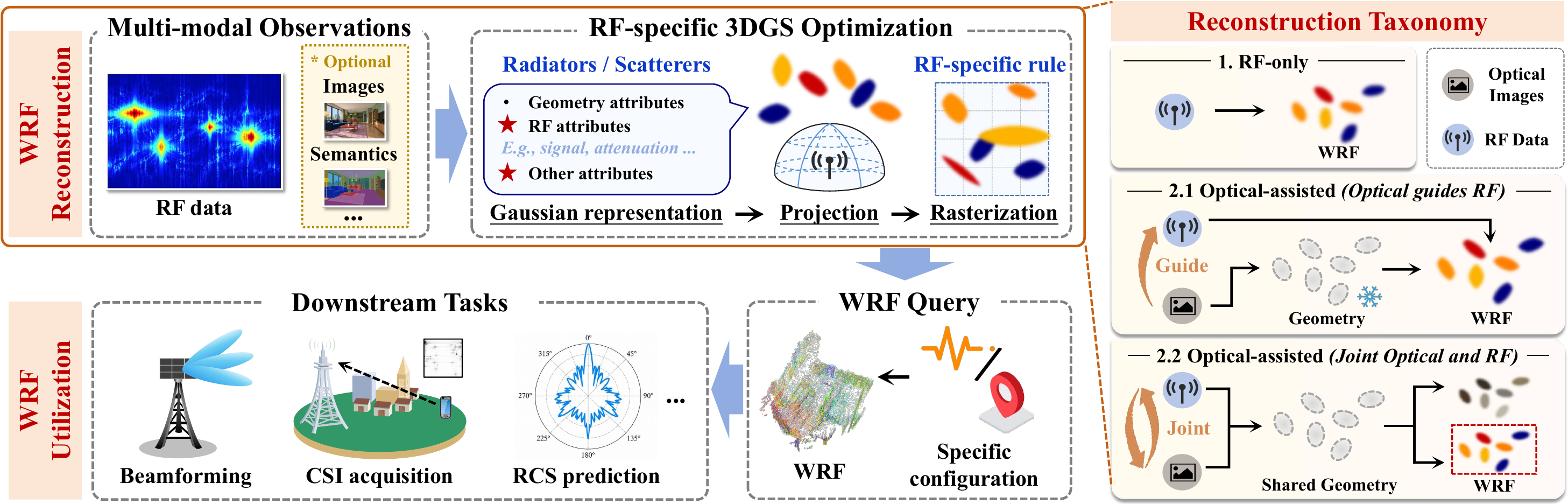}
    \caption{Overall framework and reconstruction taxonomy of 3DGS-enabled wireless environment representation. The framework comprises WRF reconstruction and utilization, with methods classified as RF-only and optical-assisted approaches. The latter further includes optical-guided and joint optical-RF strategies.}
    \label{fig:o2w_framework}
\end{figure*}

\begin{table}[!t]
\centering
\caption{Analogy between optical scene modeling and
wireless environment representation.}
\label{tab:analogy}
\renewcommand{\arraystretch}{1.3}
\begin{tabular}{ccc}
\toprule
\textbf{Aspect} & \textbf{Optical domain} & \textbf{Wireless domain} \\
\midrule
\rowcolor{LightBlue}
& Optical scene & Wireless environment \\
\rowcolor{LightBlue}
\multirow{-2}{*}{Representation} & \multicolumn{2}{c}{$\rightarrow$ \textit{Incorporate RF measurements as supervision}} \\

\multirow{2}{*}{\makecell{Gaussian\\interpretation}} 
& Scene elements & Scatterers or radiators \\
& \multicolumn{2}{c}{$\rightarrow$ \textit{Introduce RF-specific Gaussian attributes}} \\

\rowcolor{LightBlue}
& Camera image plane & RX antenna plane \\
\rowcolor{LightBlue}
\multirow{-2}{*}{\makecell{Observation\\domain}} & \multicolumn{2}{c}{$\rightarrow$ \textit{Adapt projection to RX hemispherical domain}} \\
\multirow{2}{*}{\makecell{Synthesized\\output}}
& RGB image & RF data \\
& \multicolumn{2}{c}{$\rightarrow$ \textit{Accommodate RF signal aggregation mechanisms}} \\
\bottomrule
\end{tabular}
\end{table}

In computer vision, 3DGS models how light interacts with the scene through Gaussian primitives, capturing how the resulting appearance is observed from different viewpoints. Since light and radio waves are both EM waves, they share the same governing physics, although their wavelength-dependent interactions with the environment differ. This physical connection motivates extending 3DGS from the optical domain to the wireless domain, where 3D Gaussian primitives can be adapted to characterize the interactions between radio waves and the surrounding wireless environment. Following this idea, RF data collected at multiple locations are used to reconstruct a queryable wireless environment representation, from which channel knowledge can be inferred at unmeasured locations.

Realizing this extension requires several RF-specific adaptations, as summarized in Table~\ref{tab:analogy}. In the wireless domain, Gaussians can be reinterpreted as scatterers or radiators and equipped with RF-specific attributes, such as signal and attenuation. Correspondingly, their contributions are aggregated onto the receiver (RX) antenna plane, ultimately forming the RF data observed at the RX.

\section{A New Paradigm for 3DGS-Enabled\\Wireless Environment Representation}

In this section, we first formulate a general framework for 3DGS-based wireless environment representation. We then establish a taxonomy that groups existing approaches into two representative implementation strategies and distill the key advantages of 3DGS for wireless environment representation.

\subsection{Main Framework of 3DGS for Wireless Communications}
Within the 3DGS paradigm, a wireless radiance field (WRF) composed of Gaussian primitives represents the wireless environment by characterizing the propagation and spatial distribution of radio signals. As illustrated in Fig.~\ref{fig:o2w_framework}, the general framework consists of two main stages, namely \textit{WRF reconstruction} and \textit{WRF utilization}.

The WRF reconstruction process begins with collecting observations to supervise the subsequent RF-specific 3DGS optimization. RF measurements provide essential information about wireless propagation but may offer limited environmental information. Therefore, additional modalities can provide complementary priors to facilitate WRF reconstruction.
The 3DGS pipeline generally consists of the three components introduced in Fig.~\ref{fig:3DGS}, including Gaussian representation, projection, and rasterization. However, adaptations are required to accommodate the differences between optical and radio propagation, as summarized in Table~\ref{tab:analogy}.
First, unlike Gaussian primitives in computer vision that represent light propagation, those in wireless communications characterize radio wave propagation by acting as scatterers or radiators. Accordingly, RF-specific attributes are introduced in addition to geometric attributes. Other attributes may also be incorporated depending on the available modalities. Second, the projection process needs to be adapted to the RF observation domain. Unlike pinhole cameras that project 3D scenes onto a 2D image plane, RX antenna arrays capture radio signals over a hemispherical angular domain. Third, the rendering process also requires adaptations to accommodate RF-specific Gaussian attributes and signal aggregation mechanisms. Once reconstructed, the WRF can be queried at specific configurations to predict the channel, supporting various tasks as discussed in Sec.~\ref{sec:task}.

In the following, we establish a taxonomy of existing WRF reconstruction approaches according to the adopted observation modalities. The right panel of Fig.~\ref{fig:o2w_framework} illustrates the representative reconstruction strategies.

\subsection{RF-Only Strategy}
\label{subsec:rfonly}
WRF-GS~\cite{wen2025wrf} reconstructs the WRF using RF measurements as the sole supervision modality. In WRF-GS, Gaussian primitives are interpreted as virtual radiators with RF-specific attributes, including signal and attenuation. These RF-specific attributes are predicted by an NN that takes the Gaussian centers and the transmitter (TX) position as inputs. To project the 3D Gaussians onto the RX angular domain, WRF-GS replaces perspective projection with Mercator projection, which maps spherical coordinates onto a 2D plane. The target observations are spatial spectra, which characterize the distribution of received signal power over different angles of arrival. For instance, a spatial spectrum covering a hemispherical angular domain can be represented as a $360\times90$ matrix with a one-degree angular sampling interval. Such spectra are synthesized through electromagnetic splatting, where the contributions of projected Gaussians are accumulated using signal and attenuation attributes, analogous to color and opacity in optical rendering, respectively. During training, the Gaussian geometry and RF-specific attributes are optimized using ground-truth spatial spectra as supervision.

The RF-only strategy requires no auxiliary measurements for WRF reconstruction. However, spatial spectra provide limited geometric information, potentially causing the reconstructed Gaussian primitives to deviate from the actual physical structures that influence wireless propagation. This motivates the incorporation of complementary observations to provide additional environmental information.

\begin{figure}
    \centering
    \includegraphics[width=1\linewidth]{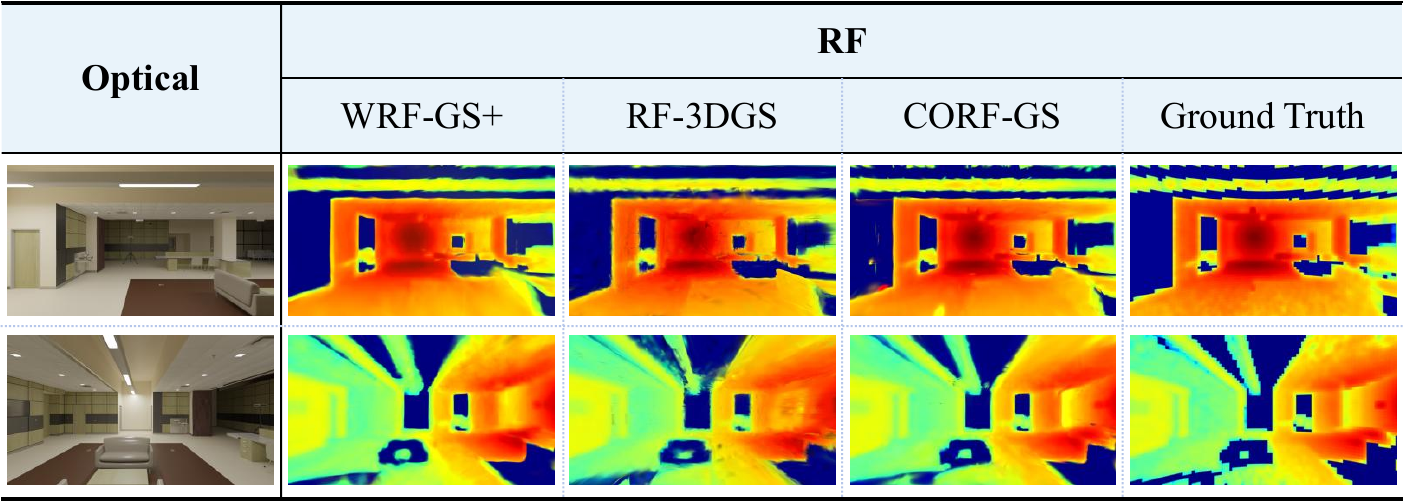}
    \caption{Spatial spectrum synthesis quality of different WRF reconstruction strategies. The correspondence between optical images and spatial spectra highlights the ability of 3DGS to capture RF propagation characteristics. WRF-GS+ extends WRF-GS~\cite{wen2025wrf} with deformable 3D Gaussians.}
    \label{fig:optical_rf}
\end{figure}

\subsection{Optical-Assisted Strategy}
Optical-assisted approaches further incorporate images to provide explicit geometric priors for WRF reconstruction. RF-3DGS~\cite{zhang2026rf} assumes that Gaussian primitives correspond to physical object surfaces that contribute to RF propagation, rather than unconstrained virtual radiators as in WRF-GS. This interpretation enables the Gaussian geometry to be constrained by the real environment. Based on this idea, RF-3DGS adopts a two-stage training pipeline. In the first stage, Gaussians are optimized using images to reconstruct the optical scene. In the second stage, the Gaussian geometry is frozen to preserve the scene structure, while the SH coefficients and opacity are re-optimized with preprocessed spatial spectra. Specifically, the SH coefficients encode view-dependent signal attributes, while the spatial spectra are transformed into ``RF images" compatible with the pinhole camera model, as shown in Fig.~\ref{fig:optical_rf}. Therefore, RF-3DGS retains the original 3DGS representation, projection, and rasterization without modifications.

Nevertheless, CORF-GS~\cite{zhang2026corf} points out that Gaussian geometry optimized solely for optical reconstruction may not adequately capture structures that are critical to wireless propagation, as light and radio waves interact differently with the environment. To address this limitation, CORF-GS allows optical images and spatial spectra to jointly optimize the shared Gaussian geometry while maintaining modality-specific appearance attributes. Specifically, Gaussians are first sampled under optical guidance and then refined using both images and spatial spectra, with a weighting factor balancing the two reconstruction objectives. As a result, the learned geometry can simultaneously capture structures relevant to both optical and wireless propagation, thereby improving WRF fidelity as shown in Fig.~\ref{fig:optical_rf}.

Overall, optical-assisted approaches introduce stronger geometric constraints and reduce the difficulty of reconstructing scene geometry and wireless responses from RF data alone. Moreover, the reconstructed scene geometry can serve as a reusable structural prior across varying TX and RX locations, potentially reducing WRF adaptation overhead.

\begin{figure}
    \centering
    \includegraphics[width=1\linewidth]{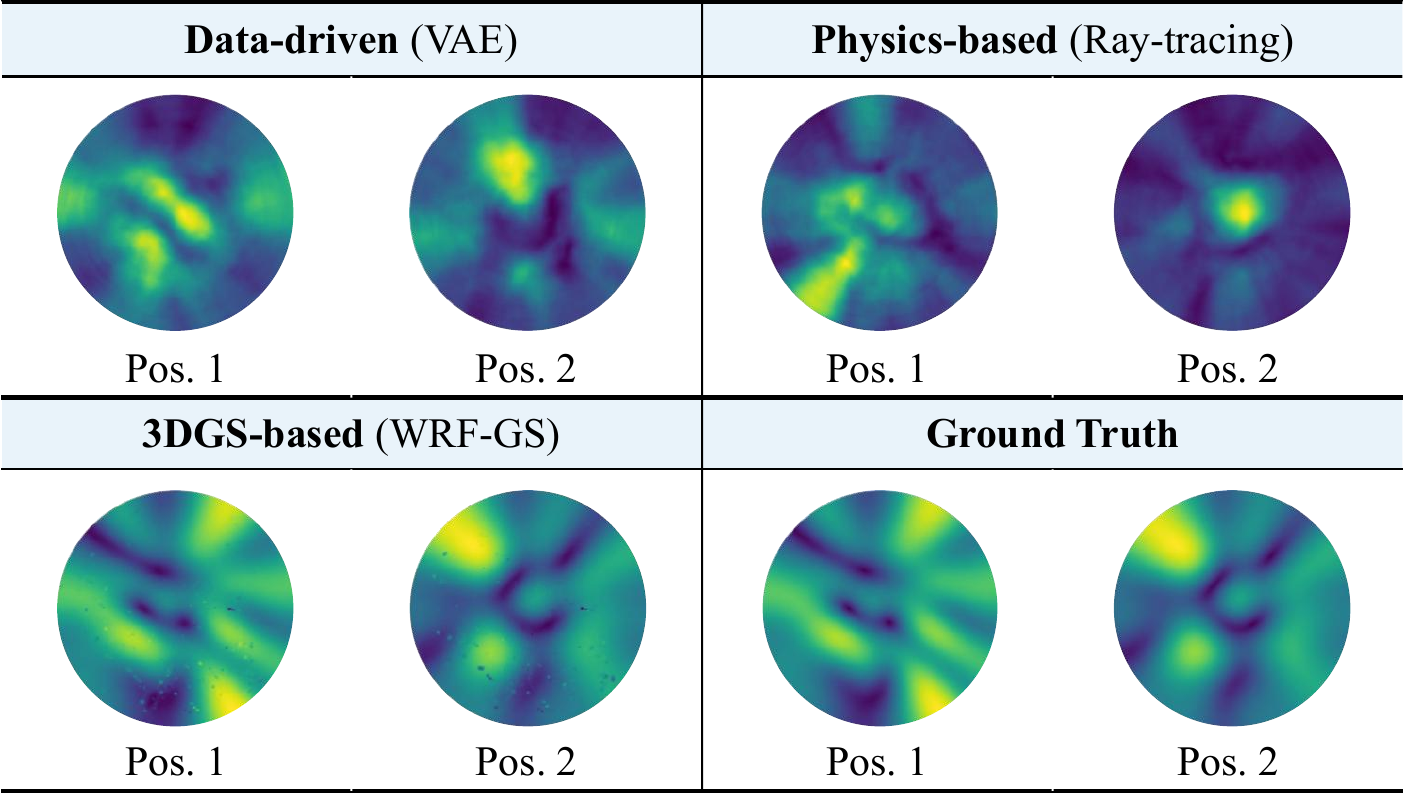}
    \caption{Qualitative comparison of spatial spectrum synthesis using data-driven, physics-based, and 3DGS-based approaches~\cite{wen2025wrf} against ground truth.}
    \label{fig:visualization}
\end{figure}

\subsection{Advantages of 3DGS for Wireless Communications}
3DGS shows great potential for wireless environment representation, as illustrated by the qualitative comparison of spatial spectrum synthesis in Fig.~\ref{fig:visualization}. The key advantages are summarized as follows.

\subsubsection{Interpretable explicit representation}
3DGS explicitly represents the wireless propagation environment using Gaussian primitives, whose RF contributions are analytically rendered into the spatial spectrum. Unlike black-box predictors, this explicit representation predicts the channel through an interpretable synthesis mechanism.

\subsubsection{Learnable environmental priors}
The wireless channel reflects the interactions between radio waves and the environment. By learning a Gaussian representation from RF data, 3DGS captures wireless environmental knowledge. Such knowledge provides reusable environmental priors for generalization enhancement and downstream sensing tasks.

\subsubsection{Differentiable rendering pipeline}
Gaussian attributes can be optimized through gradient-based learning by minimizing the loss between rendered and ground-truth RF data. Therefore, 3DGS combines an explicit wireless environment representation with the flexibility of data-driven optimization.

\subsubsection{Real-time WRF inference}
The rendering process is highly efficient owing to low-complexity projection and GPU-friendly tile-based rasterization. Therefore, a 3DGS-based WRF enables fast channel rendering at novel RX locations, which is well suited for real-time channel acquisition.

These advantages make 3DGS a promising paradigm for wireless environment representation, particularly in future environment-aware and real-time wireless systems.

\section{Downstream Applications of 3DGS-Enabled Wireless Communications}
\label{sec:task}
3DGS-based wireless environment representations encode valuable knowledge of radio propagation and the surrounding environment. In this section, we explore how such knowledge facilitates communication and sensing applications.

\subsection{Communication-Oriented Applications}
Environmental awareness can facilitate both CSI acquisition and utilization. By reconstructing an explicit WRF, 3DGS captures environmental priors, such as scatterer distributions, shared across different locations. For CSI acquisition, these priors enable channel prediction or assist channel estimation with sparse pilots, thereby reducing acquisition overhead. For CSI utilization, a WRF can render task-specific channel information, such as CSI and spatial spectra, to support beamforming, link adaptation, and symbol detection.

\subsubsection{3DGS for CSI estimation and prediction}
GeoGS-CE~\cite{zhang2026geogs} utilizes delay-beam channel priors governed by the surrounding geometry and relatively stable in high-mobility scenarios to assist channel estimation. Specifically, GeoGS-CE constructs a static 3D Gaussian map to represent non-line-of-sight (NLoS) components. Then, a deformer adapts the opacity, delay, and gain of each Gaussian based on the user position, while a line-of-sight (LoS) component is determined by TX-RX geometry. By combining NLoS and LoS contributions, a delay-beam spectrum is rendered as a covariance prior for linear minimum mean-square error estimation with sparse pilots.

Beyond CSI estimation, 3DGS also facilitates CSI prediction. In frequency-division duplex systems, although the uplink and downlink channels operate in different bands, they are governed by the same environment. Exploiting this property, WRF-GS~\cite{wen2025wrf} is extended to predict downlink CSI from uplink CSI. Based on the pipeline introduced in Sec.~\ref{subsec:rfonly}, the NN input is changed from the TX position to uplink CSI, while the rendering target is shifted from spatial spectra to downlink CSI. During training, WRF-GS captures the shared propagation characteristics underlying uplink and downlink CSI, enabling downlink CSI prediction without feedback.

\subsubsection{3DGS for beamforming}
Beamforming enhances signal quality by steering directional beams toward favorable directions. Environmental priors captured by 3DGS can help identify these directions, thereby reducing beam training overhead. Specifically, RadioSight~\cite{zhang2026radiosight} constructs a 3DGS-based WRF to render spatial spectra along predicted user trajectories. The dominant peaks indicate directions with high received signal power, which are then back-traced to the TX for predictive beamforming without instantaneous CSI acquisition.

To support continuous WRF updates in dynamic environments, RadioSight exploits optical images and semantic information to partition Gaussians into object-level clusters, distinguishing movable objects from the static background. Specifically, short-term environmental variations are handled by masking the affected regions in the rendered spatial spectrum, whereas long-term object changes are identified through semantic labels and incorporated into the WRF through object updates followed by lightweight fine-tuning.

\subsubsection{3DGS for link adaptation and detection}
GSpaRC~\cite{nukapotula2026gsparc} evaluates the utility of CSI predicted by 3DGS for system-level wireless communication tasks, including link adaptation and symbol detection. For link adaptation, the CSI predicted by GSpaRC at each RX location is converted into an effective signal-to-noise ratio, which is then mapped to an appropriate modulation and coding scheme (MCS) under a target block error rate. Compared with the optimal MCS selection based on ground-truth CSI, the 3DGS-enabled approach achieves comparable throughput, demonstrating that the predicted CSI is sufficiently accurate for practical link adaptation.

Symbol detection is further investigated through end-to-end simulations, where the conventional least-squares channel estimator is replaced by GSpaRC-based CSI prediction. During simulations, the predicted CSI is directly used for signal detection without additional pilot-based channel estimation. The results show that detection with GSpaRC-predicted CSI significantly outperforms detection with pilot-estimated CSI. Overall, these system-level evaluations highlight the capability of GSpaRC to support practical communication links by providing reliable CSI for downstream physical layer tasks.

\begin{figure*}
    \centering
    \includegraphics[width=0.9\linewidth]{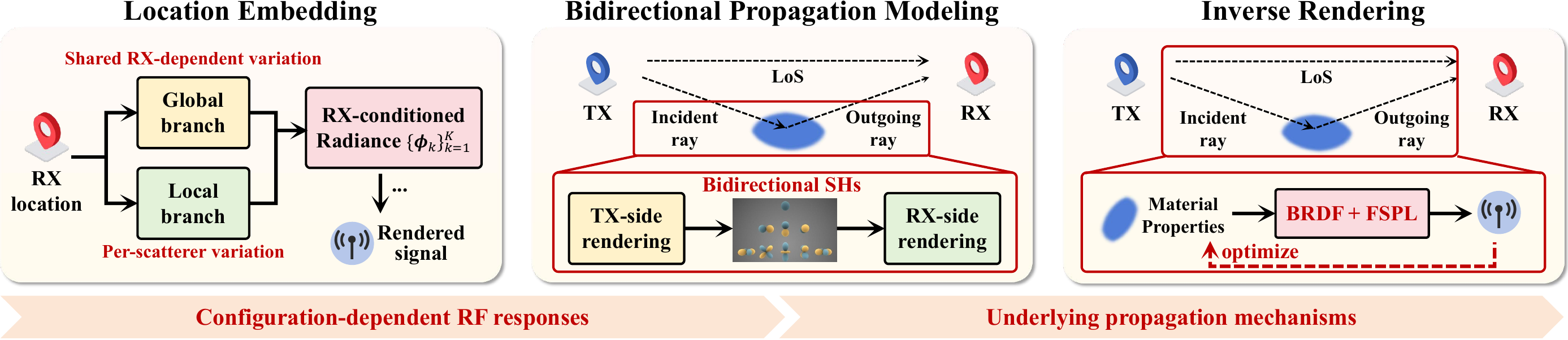}
    \caption{Three strategies for enhancing generalization to unseen TX/RX locations, including location embedding, bidirectional propagation modeling, and inverse rendering. These strategies illustrate a shift from learning configuration-dependent RF responses toward modeling propagation mechanisms.}
    \label{fig:unseentr}
\end{figure*}

\subsection{Sensing-Oriented Applications}
Radio propagation is shaped by the surrounding environment, leaving signatures of scene properties and object states in received RF signals. By reconstructing a WRF, 3DGS encodes such environmental information into Gaussian attributes, supporting sensing tasks such as scene reconstruction, radar cross-section (RCS) prediction, and object detection.

\subsubsection{Scene reconstruction}
Spatial spectra depend on both Gaussian geometry and RF appearance, with the latter influenced by TX locations and material properties. Therefore, URF-GS~\cite{wen2026bridging} introduces material attributes into Gaussian primitives, such as roughness and metallicity. Similar to RF-3DGS, URF-GS adopts a two-stage training pipeline. However, in the second stage, URF-GS optimizes material attributes rather than SH coefficients and opacity. To render spectra, URF-GS employs a bidirectional reflectance distribution function (BRDF) to model material-dependent reflection and scattering as functions of incident and outgoing directions, while incorporating free-space path loss (FSPL) to account for propagation attenuation. In this inverse rendering process, URF-GS reconstructs the scene by learning material attributes from spatial spectra.

\subsubsection{RCS prediction}
RCS characterizes how objects reflect radio waves and is closely related to their geometry and EM properties. RFIR~\cite{wang2026radio} evaluates RCS synthesis performance across diverse objects, viewpoints, frequencies, and distances. Similar to URF-GS, RFIR reconstructs the scene through inverse rendering. Each Gaussian primitive is endowed with surface normals, material EM parameters, and roughness. It further incorporates frequency-dependent modeling to support RCS synthesis at 2.4 and 5.8~GHz as well as wideband settings. Experiments on both simulated and measured data demonstrate that the recovered scene representation can accurately reproduce RCS patterns across varying configurations.

\subsubsection{Object detection}
RF-3DGS~\cite{zhang2026rf} demonstrates that a previously reconstructed WRF can serve as a background propagation prior. In conventional sensing, scattering caused by an object is often mixed with background multipath, making it difficult to isolate the response of the object of interest. To address this, RF-3DGS first reconstructs a WRF and renders a prior spatial spectrum at an RX location of interest. When a new object enters the scene, the real-time wireless feedback is compared with the prior spectrum to identify newly introduced propagation components. The difference reveals the presence of the object and can further support localization or recognition.

\section{Practical Deployment of 3DGS-Enabled Wireless Communications}
Moving from simulations toward real-world 3DGS-enabled wireless systems faces several challenges. This section investigates the main concerns for practical deployment, including generalization enhancement, real-time channel representation and prediction, and data generation and validation.

\subsection{Generalization Enhancement}
Wireless channels vary with TX/RX locations and operating frequencies, while 3DGS-based WRFs are often optimized for fixed settings, requiring costly retraining to accommodate such variations. To address this limitation, strategies for generalization to unseen locations and frequencies are explored.

\subsubsection{Unseen locations}
\label{subsubsec:loc} 
Existing studies mainly adopt three strategies to enable WRF generalization to unseen TX/RX locations, as illustrated in Fig.~\ref{fig:unseentr}. The most straightforward solution is to condition the WRF on the location to be generalized. For example, RxGS~\cite{yang2026rxgs} incorporates the RX location into the architecture to predict RX-conditioned radiance.
BiWGS~\cite{zhou20256d} further models bidirectional propagation to generalize across different TX-RX configurations. Instead of using vanilla SHs that depend on a single direction, it employs bidirectional SHs to characterize each Gaussian response over both incident and outgoing directions. URF-GS~\cite{wen2026bridging} adopts a more physics-informed strategy through inverse rendering. It infers persistent material properties from spatial spectra and reuses them for forward rendering under new TX-RX configurations. These developments indicate a shift from directly learning configuration-dependent RF responses toward modeling the underlying propagation mechanisms.

\subsubsection{Unseen frequencies}
To enable cross-band channel modeling, XBandGS~\cite{li2026cross} employs shared Gaussian geometry across frequency bands and introduces a frequency-conditioned mixture-of-experts (MoE) architecture to model Gaussian RF attributes, including signal and attenuation. Specifically, the MoE consists of a gating NN and four experts, each specializing in LoS, specular reflection, diffuse scattering, and diffraction, respectively. The gating NN takes frequency embeddings and geometric features derived from TX-RX positions and Gaussian geometry as inputs to assign adaptive weights to the experts. Each expert incorporates a mechanism-specific analytical propagation model, such as free-space path loss or Fresnel reflection, to characterize frequency-dependent RF contributions. The expert outputs are then weighted to form the Gaussian RF attributes, enabling spatial spectrum synthesis at unseen frequencies.

\subsection{Real-Time Channel Representation and Prediction}
Timely and reliable channel information is critical in dynamic wireless environments. However, offline WRFs may become outdated as environments evolve, while excessive Gaussian primitives increase rendering latency. Therefore, online adaptation and lightweight rendering are essential for real-time wireless communication.

\subsubsection{Online adaptation}
CORF-GS~\cite{zhang2026corf} incrementally reconstructs the WRF from sequential optical-RF observations, each containing an image, a spatial spectrum, and pose information. As each new observation arrives, the existing WRF is progressively expanded and refined using both optical and RF data. Instead of relying on gradient-based Gaussian expansion, CORF-GS directly samples new primitives in under-represented regions guided by optical structures. Simulation results show that CORF-GS improves the peak signal-to-noise ratio by 1.6 dB over RF-3DGS, while reconstructing the WRF from 128 training observations in 2.30 minutes, equivalent to approximately 1.1 seconds of optimization per observation.

\subsubsection{Lightweight rendering}
3DGS achieves fast rendering by adopting rasterization and explicit Gaussian primitives. However, densification may introduce redundant Gaussians, leading to overparameterization, increased storage, and reduced rendering efficiency. Inspired by NN pruning, P-WRFGS~\cite{bian2026pruning} assigns a learnable importance mask to each Gaussian and introduces a regularization term to balance rendering quality and efficiency. Low-importance Gaussians are then adaptively pruned during training, reducing storage and rendering costs.

\subsection{Data Generation and Validation}
Real-world RF data collection is costly and time-consuming, posing challenges for algorithm development and evaluation. As a novel view synthesis technique, 3DGS enables efficient data generation, while real-world validation remains essential to assess the accuracy and reliability of reconstructed WRFs.

\subsubsection{Data generation}
HybridSim~\cite{xiong2026hybridsim} synthesizes high-fidelity mmWave radar signals for dynamic human sensing by decoupling wireless propagation into direct and indirect components. The direct path models single-bounce reflections through inverse rendering, whereas the indirect path represents multi-bounce reflections using 3DGS. This decoupling mitigates the entanglement between dynamic human reflections and environmental multipath, helping preserve the micro-Doppler signatures essential for human sensing. Moreover, modeling complex multipath using 3DGS avoids costly ray-tracing, enabling efficient and high-fidelity data synthesis for downstream sensing tasks.

\subsubsection{Real-world validation}
The sim-to-real gap makes real-world validation essential for practical deployment. An initial step is to calibrate a simulated WRF using physical measurements. For instance, RF-3DGS~\cite{zhang2026rf} uses 60-GHz field measurements collected by a channel sounder and processed with the space-alternating generalized expectation-maximization algorithm to fine-tune a WRF pretrained on simulated data. RadioSight~\cite{zhang2026radiosight} takes a further step by implementing a dynamic 3DGS-based WRF on a full-stack 28-GHz prototype and evaluating it in practical beam prediction and handover tasks. These system-level evaluations directly demonstrate the practical utility of 3DGS in wireless environment representation.

\section{Open Issues and Future Directions}
Although 3DGS-enabled wireless communications have shown promising performance, several challenges remain to be addressed. To further advance this paradigm, key open issues and future research directions are summarized below.

\subsection{From Scene-Specific to Cross-Scene WRF Reconstruction}
Cross-scene generalization is essential for large-scale deployment of 3DGS-based WRFs, since reconstructing a separate model for each scene would incur substantial data collection and computational costs. Optimization-based 3DGS~\cite{kerbl20233d} directly optimizes scene-specific Gaussian primitives and requires additional training for each new environment. In contrast, feed-forward 3DGS trains NNs to predict Gaussian primitives from input observations. Inspired by this, future WRFs could learn generalizable mappings from RF measurements across diverse scenes to Gaussian primitives, enabling fast reconstruction in unseen scenes. Foundation models and generative models may further enhance cross-scene generalization.

\subsection{Channel Modeling for XL-MIMO}
The emergence of extremely large-scale MIMO (XL-MIMO) significantly enlarges the array aperture, shifting the channel from conventional far-field models to near-field propagation characterized by spherical wavefronts and spatial non-stationarity. While existing 3DGS-based WRFs are predominantly tailored to far-field assumptions, their applicability to near-field XL-MIMO channels remains largely unexplored. Extending wireless environment representations to accurately capture these near-field characteristics is therefore an important research direction. To bridge this gap, adapting the 3DGS framework with near-field projection and spherical-wavefront rendering mechanisms is a possible solution.

\subsection{Real-World Data Collection}
Acquiring high-quality RF data for WRF reconstruction remains a key challenge in the practical deployment of 3DGS-enabled wireless systems. Reliable channel measurements require hardware calibration, while the angular and delay resolutions of RF observations are constrained by antenna aperture and bandwidth, respectively. Moreover, collecting RF data across various locations incurs substantial overhead, motivating efficient measurement strategies that balance acquisition cost and spatial coverage. Multi-modal data collection introduces additional challenges in spatial and temporal alignment due to differences in sensing perspectives, spatial resolutions, and sampling rates. Therefore, improving the quality and efficiency of real-world data collection remains a vital direction.

\subsection{WRF Lifecycle Management for Practical Deployment}
Reliable deployment of 3DGS-based WRFs in dynamic environments requires lifecycle management covering reconstruction, performance monitoring, adaptive updates, and version management. Specifically, WRF validity can be assessed through downstream task performance monitoring. This assessment guides adaptive update policies that set update triggers, frequency, and scope to prevent severe degradation with minimal overhead. Moreover, cross-device WRF sharing for cooperative tasks poses version consistency challenges. Overall, future research should establish standardized lifecycle management for reliable long-term operation.

\section{Conclusion}
In this article, we provide a systematic study of 3DGS as a new paradigm for wireless environment representation, highlighting its ability to learn complex RF characteristics under explicit geometric constraints. We first clarify the rationale for extending 3DGS from visual scene modeling to wireless communications. Then, a general framework for 3DGS-based wireless environment representation is established, based on which its downstream applications and key considerations for deployment are further examined. Several promising directions are also highlighted, including cross-scene generalization, near-field channel modeling, real-world data collection, and WRF lifecycle management.

\bibliographystyle{IEEEtran}
\bibliography{IEEEabrv,reference}
\vspace{-1cm}

\begin{IEEEbiographynophoto} {Jinya Zhang}
[Student Member, IEEE] (zhangjinya@seu.edu.cn) is currently
working towards her Ph.D. degree with Southeast University.
\end{IEEEbiographynophoto}
\vspace{-1cm}
\begin{IEEEbiographynophoto} {Jiajia Guo} 
[Member, IEEE] (jiajiaguo@seu.edu.cn) received the Ph.D. degree from Southeast University in 2023 and focuses on AI-native air interfaces.
\end{IEEEbiographynophoto}
\vspace{-1cm}
\begin{IEEEbiographynophoto} {Chao-Kai Wen} [Fellow, IEEE] (chaokai.wen@mail.nsysu.edu.tw) received the Ph.D. degree from National Tsing Hua University in 2004. He is currently a Professor with National Sun Yat-sen University. His research interests center around the optimization of wireless multimedia networks.
\end{IEEEbiographynophoto}
\vspace{-1cm}
\begin{IEEEbiographynophoto} {Shi Jin} [Fellow, IEEE] (jinshi@seu.edu.cn) received his Ph.D. degree in information and communications engineering from Southeast University in 2007. He is currently with Southeast University. His research interests include wireless communications, random matrix theory, and information theory.
\end{IEEEbiographynophoto}

\end{document}